\documentstyle[12pt]{article}
\def\baselinestretch{1.5}
\newcommand{\reals}{\mbox{${\rm I\!R }$}}

\newcommand{\ant}{\int\limits}

\newcommand{\cam}{{\cal M}}
\newcommand{\caz}{{\cal Z}}

\newcommand{\caf}{{\cal F}}

\newcommand{\beq}{\begin{eqnarray}}
\newcommand{\eeq}{\end{eqnarray}}
\newcommand{\nn}{\nonumber}
\newcommand{\mnintx}{\ant_{\cam}d^nx|g(x)|^{\frac 1 2}}
\newcommand{\mvintx}{\ant_{\cam}d^4x|g(x)|^{\frac 1 2}}
\newcommand{\mninty}{\ant_{\cam}d^ny|g(y)|^{\frac 1 2}}

\newcommand{\back}{\hat{\phi}}
\newcommand{\abl}{\partial}
\newcommand{\tintnu}{\ant_0^{\infty}dt}

\newcommand{\mnrs}{\mu\nu\rho\sigma}
\newcommand{\mn}{\mu\nu}

\newcommand{\q}{\left(\xi -\frac 1 6\right)}

\newcommand{\vp}{(4\pi)^2}
\newcommand{\ep}{\epsilon}

\renewcommand{\theequation}{\mbox{\arabic{section}.\arabic{equation}}}

\begin{document}
\def\bce{\begin{center}}
\def\ece{\end{center}}
\def\bea{\begin{eqnarray}}
\def\eea{\end{eqnarray}}
\def\brr{\begin{array}}
\def\err{\end{array}}
\def\ben{\begin{enumerate}}
\def\een{\end{enumerate}}
\def\bei{\begin{itemize}}
\def\eei{\end{itemize}}
\def\ul{\underline}
\def\ni{\noindent}
\def\bs{\bigskip}
\def\ms{\medskip}
\def\dsp{\displaystyle}
\def\wt{\widetilde}
\def\wh{\widehat}
\def\rl{$\Real$}
\def\tu{\bigtriangleup}
\def\td{\bigtriangledown}

\hfill UB-ECM-PF 94/1

\hfill January 1994

\vspace*{3mm}

\begin{center}

{\LARGE \bf
Effective Lagrangian and the back-reaction problem in a
self-interacting $O(N)$ scalar theory \\in curved spacetime}

\vspace{4mm}

\renewcommand
\baselinestretch{0.8}
\ms

\renewcommand
\baselinestretch{1.4}
{\sc E. Elizalde}\footnote{E-mail: eli@ebubecm1.bitnet,
eli@zeta.ecm.ub.es} \\ Department E.C.M. and I.F.A.E., Faculty
of Physics,
University of \\Barcelona, Diagonal 647, 08028 Barcelona, \\
and Center for Advanced Studies, C.S.I.C., Cam\'{\i} de Santa
B\`arbara, \\
17300 Blanes, Catalonia, Spain,\\
{\sc K. Kirsten}\footnote{E-mail: klaus@ebubecm1.bitnet. Alexander von
Humboldt Foundation Fellow.}

 and {\sc S.D. Odintsov}\footnote{E-mail: odintsov@ebubecm1.bitnet.
On leave from: Tomsk Pedagogical Institute, 634041 Tomsk, Russian
Federation.} \\  Department E.C.M., Faculty of Physics,
University of  Barcelona, \\  Diagonal 647, 08028 Barcelona,
Catalonia,
Spain \\

\renewcommand
\baselinestretch{1.4}

\vspace{5mm}

{\bf Abstract}

\end{center}

A derivation of the one-loop effective Lagrangian in the
self-interacting
$O(N)$ scalar theory, in slowly varying gravitational fields, is
presented (using $\zeta$-regularization and heat-kernel techniques).
The result is given in terms of the expansion in powers of the curvature
tensors (up to quadratic terms) and their derivatives, as well as in
derivatives of the background scalar field (up to second derivatives).
The
renormalization group improved effective Lagrangian is studied,
what gives the leading-log approach of the whole perturbation theory.
An analysis of the effective equations (back-reaction problem) on the
static hyperbolic spacetime $\reals^2 \times H^2/\Gamma$ is carried out
for
the simplest version of the theory: $m^2=0$ and $N=1$. The existence of
the solution $\reals^2 \times H^2/\Gamma$, induced by purely quantum
effects, is shown.

\vspace{4mm}


\newpage
\section{Introduction}
\setcounter{equation}{0}
There is little to say about the importance of the effective
action
formalism in modern quantum field theory and quantum cosmology. If
we had at our disposal a closed form for the action of the
theory of quantum
gravity ---which still does not exist as a consistent theory--- we would
be able to construct a reasonable picture of the early universe which,
presumably, would contain the inflationary stage (for a review see
\cite{kolbturner90,linde90,brandenberger85}).
Unfortunately, even in frames of the rather simple approximations to
quantum
gravity based on quantum field theory in curved spacetime (for a review
see \cite{dewitt65a,birelldavies82,buchbinderodintsovshapiro92}),
it is impossible to find the effective action in a closed form.

First of all, one has to work in perturbation theory and, as a rule, in
the
simplest one-loop approximation. However, even in this case there appear
some limitations, connected with the fact that it is impossible
to calculate the one-loop effective action in general curved spacetime.

For that reason,
there are a lot of explicit calculations of one-loop
effective actions mainly for some simple models (like the $\lambda
\varphi^4$-theory and scalar electrodynamics) in specific
backgrounds paying special attention to the role of constant curvature
[7]-[11],
\nocite{shore80a,oconnorhushen83,allen83,vilenkin83,buchbinderodintsov89a}
topology [12]-[22],
\nocite{toms82,fordyoshimura79,ford80,toms80,fordtoms82,denardospallucci80}
\nocite{denardospallucci80a,actor90,elizalderomeo90a,elizaldekirsten93unv}
\nocite{hosotani84}
a combination of both
\cite{kennedy81,cognolakirstenzerbini93,bytsenkokirstenodintsov93},
and, finally, to the anisotropy in different Bianchi-type universes
[26]-[33].
\nocite{critchleydowker82,oconnorhushen85,oconnorhu86,futamase84}
\nocite{ringwald87,ringwald87a,hartlehu79,berkin92}

If we still want to obtain information connected with general
properties of curved spacetime, a very useful method is the
quasilocal approximation scheme for the effective action in curved
spacetime [6], [34]-[37].
\nocite{buchbinderodintsov85,buchbinderodintsovshapiro92}
\nocite{oconnorhu84,critchleyhustylianopoulos87}
\nocite{cognolakirstenvanzo93}
This will be the main method applied in the present work,
where we continue the study of the effective Lagrangian in curved
spacetime. In particular, using heat-kernel techniques, we will derive
the one-loop effective Lagrangian in a scalar self-interacting
$O(N)$-theory,
thus generalizing existing results for the $\lambda \varphi^4$
theory in curved spacetime.
(For the $O(N)$-sigma model in two dimensions see
\cite{wiesendangerwipfunv}).

Some technical problems, which appear in such calculations are
discussed in sections 2 and 3, where the computation of the one-loop
effective action and its renormalization is performed.
Section 4 is devoted to an attempt to discuss higher-loop effects.
For that reason we use the renormalization group (RG) to
improve the effective Lagrangian and to obtain it in the leading-log
approximation of the whole perturbation theory.
In section 5 we analyse the back-reaction problem, following the
early attempts of [32], [39]-[42]
\nocite{horowitzwald78,fischettihartlehu79,hartlehu79,hartlehu80,anderson83}
(for a review see \cite{birelldavies82}), using the static hyperbolic
space $\reals^2 \times H^2/\Gamma$ with varying radius as a background,
and
using two different approximations for the effective action: the exact
one-loop effective action \cite{bytsenkokirstenodintsov93} and the RG
improved effective action. The existence of a quantum solution induced
by matter is shown.
Section 6 summarizes the main
results. Finally, appendix A presents the technique of diagonalization
of the
matrix describing the potential of the $O(N)$-field, which we need for
the calculation of the one-loop effective action.

\section{The one-loop effective action}
\setcounter{equation}{0}
The aim of this section is to derive a quasilocal approximation for the
effective action in a self-interacting $O(N)$-symmetric model described
by the action
\beq
S[\tilde{\phi},g_{\mu\nu}]=-\mnintx\left[\frac 1 2
\tilde{\phi}\Delta\tilde{\phi}-\tilde U (\tilde{\phi})\right].\label{21}
\eeq
Here $\cam$ is a smooth $n$-dimensional manifold with Lorentzian metric
$g_{\mu\nu}=diag(-,+,...,+)$,
$\tilde{\phi}=(\tilde{\phi_1},...,\tilde{\phi_N})$ is an $N$-component
scalar field, and $ \tilde U (\tilde{\phi})$ is a potential describing
the self-interaction of the scalar fields and which contains,
furthermore,
local expressions of dimension $n$, involving curvature tensors and
nonquadratic terms in the field, independent up to a total divergence.
The latter terms have to be included to ensure renormalizability of the
theory.

Expanding the action (\ref{21}) around its classical minimum $\back$ in
powers of the fluctuations $\phi =\tilde{\phi}-\back$, leads to
\beq
S[\tilde{\phi},g_{\mu\nu}]=S^{(0)}+S^{(1)}+S^{(2)}+...\nn
\eeq
Here $S^{(0)}$ describes the contribution of the classical background
field $\back$, $S^{(0)}=S[\back , g_{\mu\nu}]$, $S^{(1)}$ is linear in $
\phi$ and
\beq
S^{(2)}=\frac 1 2 \mnintx\mninty \phi_i(x) \cam_{ij}\phi_j(y)\nn
\eeq
contains the relevant quadratic contributions of the fluctuations, the
only relevant ones in the one-loop calculation we are going to perform.
Obviously,
\beq
\cam_{ij}=\frac{\delta^2S}{\delta\phi(x)\delta\phi(y)}
\left|_{\phi_i=\back_i},\right.\nn
\eeq
leading to the fluctuation operator
\beq
D=-\Delta+U(\back),\label{22}
\eeq
with
\beq
U(\back )_{ij}=\tilde U ''(\back)e_ie_j+\frac{\tilde U '(\back )}
{\back}(\delta_{ij}-e_ie_j),\label{23}
\eeq
where the normalized background field $e_i =\back_i/\sqrt{\back^2}$ has
been used.

The effective action of the theory is then expanded in powers of $\hbar$
as
\beq
\Gamma [\back] =S[\back , g_{\mu\nu}]+\Gamma ^{(1)}+\Gamma
'\,\, ,\label{24} \eeq
with the one-loop contribution $\Gamma^{(1)}$ to the action being
\beq
\Gamma^{(1)}= \frac i 2 \ln\det \frac D {\mu^2}\label{25}
\eeq
and higher-loop quantum corrections $\Gamma '$. The introduction of the
arbitrary mass parameter $\mu$ is necessary in order to keep the action
dimensionless.

The most interesting quantity for us in this section will be the
derivative expansion of the effective action (\ref{24}), that is
\beq
\Gamma [\back ,g_{\mu\nu}] =\mnintx & &\left[V(\back)+\frac 1 2
\caz_1(\back) (\back \Delta \back )+\frac 1 2 \caz_2(\back)\Delta
\back^2 \right.\nn \\
& &\left.+\frac 1 2 \caz_3 (\back)\frac{\Delta
\back^4}{\back^2}+...\right],\label{26}
\eeq
where as many as possible of the derivative terms are assumed to
be written in a form depending only on $O(N)$-symmetric quantities.

Our first goal will be to compute the quantities $V(\back )$ and
$\caz_i(\back)$ in equation (\ref{26}) up to quadratic powers in the
curvature tensors
and up to
second derivatives of the curvature terms. The tool we shall use to
achieve this result
will be zeta-function regularization in combination with
heat-kernel techniques. In the zeta-function regularization
scheme the functional determinant in eq.~(\ref{25}) is defined by
\cite{hawking77,critchleydowker76}
\begin{eqnarray}
\Gamma^{(1)}[\hat{\phi}]=-\frac i2\left[\zeta'_{D}(0)-\zeta_D(0)\ln\mu^2
\right],
\label{27}
\end{eqnarray}
where $\zeta_D (s)$ is the zeta-function associated with the operator
$D$, eq.~(\ref{22}), and the prime denotes differentiation with
respect to $s$. This means that, $\lambda_j$
being the eigenvalues of $D$, the zeta-function $\zeta_D(s)$ is defined
by
\begin{eqnarray}
\zeta_D (s)&=&\sum_j \lambda_j^{-s}
= \frac {i^s} {\Gamma(s)} \sum_j \int_0^{\infty}dt
\,\,t^{s-1}e^{-i\lambda_j t}\nonumber\\
&=&\frac {i^s}{\Gamma(s)}
\int_0^{\infty}dt\,\,t^{s-1} \,\mbox{tr}\, K(x,x,t),
\label{28}
\end{eqnarray}
where the kernel $K(x,x',t)$ satisfies the equations
\begin{eqnarray}
i\frac {\partial}{\partial t} K(x,x',t)&=&DK(x,x',t),\nonumber\\
\lim_{t\to 0}K(x,x',t) &=& |g|^{-\frac 1 2} \delta (x,x').\label{29}
\end{eqnarray}
In order to obtain the derivative expansion, eq.~(\ref{26}), of
the effective action, the following ansatz by Parker and Toms is
suggested \cite{parkertoms85,parkertoms85a}
\begin{eqnarray}
K(x,x',t)=-i\frac{\Delta_{VM}(x,x')}{(4\pi it)^{n/2}}
\;\Omega(x,x',t)\;
\exp\left\{i\left[\frac{\sigma^2(x,x')}{4t}-t\left(U(\back)-\frac 1 6 R
\right)\right]\right\},
\phantom{aa}\label{210}
\end{eqnarray}
where $\sigma(x,x')$ is the proper arc length along
the geodesic $x'$ to $x$ and $\Delta_{VM}(x,x')$ is the Van
Vleck-Morette determinant.
For $t\to 0$ the function $\Omega(x,x',t)$
may be expanded in an asymptotic series
\begin{eqnarray}
\Omega (x,x',t) =\sum_{l=0}^{\infty}a_l(x,x') (it)^l,\label{211}
\end{eqnarray}
where the coefficients $a_l$ have to fulfill some recurrence relations.
Using the ansatz (\ref{210}), it has been shown in
ref.~\cite{jackparker85},
that the dependence of $a_l$, $l=1,...,\infty$,
on the field $\hat{\phi}$ is only through derivatives of the field.

Up to the order to which we are going to calculate the effective action,
we
need only to include contributions up to $a_3$. Then, for the zeta
function eq.~(\ref{28}) one finds
\beq
\zeta_D(s)&=&-\frac{i i^s}{\Gamma(s)(4\pi i)^{\frac n 2}}\tintnu\,\,
      t^{s-1-\frac n 2}\label{212}\\
& &\times\mbox{tr}\left\{\exp\left[-it\left(U(\back ) -\frac 1 6
R\right)\right]\left[a_0-a_2t^2-ia_3t^3+...\right]\right\},\nn
\eeq
where $a_1=0$ has been used.
In order to calculate the trace in equation (\ref{212}), one has to
diagonalize the matrix $U(\back )-\frac 1 6 R$. This is done in Appendix
A.

For the calculation of the effective action let us restrict ourselves to
the
most interesting case $\tilde U (\back)=(1/2)(m^2+\xi R)\back^2+(\lambda
/4!)\back^4$ in $n=4$ dimensions. Then, as is seen by diagonalizing the
matrix $U(\back )$, two different masses
\beq
m_1^2&=&m^2+\q R+\frac {\lambda} 2 \back^2,\nn\\
m_2^2&=&m^2+\q R+\frac {\lambda} 6 \back^2,\nn
\eeq
arise. By denoting $\Gamma_i^{(1)}$ the part of the effective action
that results from the coefficient $a_i$ and by using
\beq
b_2=\frac 1 {180} (\Delta R+R_{\mnrs}R^{\mnrs}-R_{\mn}R^{\mn})\nn
\eeq
for the purely geometrical part of the $a_2$-coefficient, we find
\beq
\Gamma_0^{(1)}&=&\frac 1 {32\pi^2} \mvintx \left\{\frac {m_1^4} 2
\left[\ln\left(\frac{m_1^2}{\mu^2}\right)-\frac 3
2\right]\right.\label{213}\\
& &\left.\qquad +(N-1)\frac {m_2^4} 2
     \left[\ln\left(\frac{m_2^2}{\mu^2}\right)-\frac 3
2\right]\right\},\nn\\
\Gamma_2^{(1)}&=&\frac 1 {32\pi^2} \mvintx\nn\\
& &\times\left\{\frac{\lambda}{18}\ln\left(\frac{m_1^2}{m_2^2}\right)
     \left[-2\back\Delta\back+\frac 1 2 \Delta \back^2-\frac 1
{4\back^2}\Delta \back^4\right]\right.\nn\\
& &+\ln\left(\frac{m_1^2}{m_2^2}\right)\left[b_2-\frac 1 6 \Delta
m_2^2\right]\label{214}\\
& &\left.+\ln\left(\frac{m_2^2}{\mu^2}\right)\left[Nb_2-\frac 1 6
(N-1)\Delta m_2^2-\frac 1 6 \Delta m_1^2\right]\right\},\nn\\
\Gamma_3^{(1)}&=&\frac 1 {32\pi^2} \mvintx \label{215}\\
& &\times
\left\{-\frac{\lambda^2}{108m_1^2}\left[-\back^2(\back\Delta\back)-\frac
1 2 \back^2\Delta\back^2+\frac 1 2 \Delta \back^4\right]\right.\nn\\
&
&\left.-\frac{\lambda^2}{108m_2^2}\left[-\back^2(\back\Delta\back)-\frac
{N-4} 4 \back^2\Delta\back^2+\frac {N-2} 8 \Delta
\back^4\right]\right\}.\nn
\eeq
The complete one-loop effective action is then given as the sum
\beq
\Gamma^{(1)}=\Gamma_0^{(1)}+\Gamma_2^{(1)}+\Gamma_3^{(1)}\label{216}
\eeq
of the contributions in equations (\ref{213})-(\ref{215}).
\section{Renormalization}
\setcounter{equation}{0}
The quantum correction $\Gamma ^{(1)}[\hat{\phi}]$ depends on the
arbitrary
renormalization scale $\mu$. This dependence will be now removed by the
renormalization procedure which we are going to describe.
As is well known by now, one is always forced to take into consideration
the most
general quadratic gravitational Lagrangian \cite{utiyamawitt62}, and so
one is led to consider the classical Lagrangian
\begin{eqnarray}
L_{cl}&=&\eta\Box\hat{\phi}^2
-\frac 1 2 \hat{\phi} \Box \hat{\phi} +\Lambda +\frac 1 {24}
\lambda \hat{\phi} ^4
+\frac 1 2 m^2 \hat{\phi} ^2 +\frac 1 2 \xi R \hat{\phi}^2 \nonumber\\
&&+\varepsilon_0 R+\frac 1 2 \varepsilon_1 R^2 +\varepsilon_2 C
+\varepsilon_3 G +\varepsilon_4 \Box R,
\label{217}
\end{eqnarray}
with the corresponding counterterm contributions
\begin{eqnarray}
\delta L_{cl}&=&\delta\eta\Box\hat{\phi}^2
+\delta\Lambda+\frac 1 {24} \delta\lambda \hat{\phi} ^4
+\frac 1 2 \delta m^2 \hat{\phi} ^2 +\frac 1 2 \delta\xi R \hat{\phi}^2
 \nonumber\\
&&+\delta\varepsilon_0 R+\frac 1 2 \delta\varepsilon_1 R^2
 +\delta\varepsilon_2 C
+\delta\varepsilon_3 G +\delta\varepsilon_4 \Box R
\label{218}
\end{eqnarray}
which are necessary in order to renormalize all coupling constants.
With $C$ and $G$ we indicate, respectively, the square of the Weyl
tensor and the Gauss-Bonnet density. They read
\begin{eqnarray}
C&=&R_{\mu\nu\rho\sigma}R^{\mu\nu\rho\sigma}
-2R_{\mu\nu}R^{\mu\nu}+\frac13R^2,\\
G&=&R_{\mu\nu\rho\sigma}R^{\mu\nu\rho\sigma}
-4R_{\mu\nu}R^{\mu\nu}+R^2.
\end{eqnarray}
The renormalization conditions are given by \cite{oconnorhu84}
\begin{eqnarray}
\Lambda &=&L\left|_{\hat{\phi} =\varphi_0,R=0},\right.\nonumber\\
\lambda &=&\frac{\partial^4 L}{\partial \hat{\phi}^4}\left|_{\hat{\phi}
=\varphi
      _1,R=0},\right.\nonumber\\
m^2 &=&\frac{\partial ^2 L}{\partial \hat{\phi}^2}\left|_{\hat{\phi} =0,
          R=0},\right.\nonumber\\
\xi &=&\frac{\partial L}{\partial R \partial
\hat{\phi}^2} \left|_{\hat{\phi} =\varphi_3,
               R=R_3},\right.\nonumber\\
\varepsilon_0 &=&\frac{\partial L}{\partial R}
\left|_{\hat{\phi} =0, R=0},\right.\label{34}\\
\varepsilon_ 1&=& \frac{\partial ^2 L}{\partial R^2}
\left|_{\hat{\phi} =0, R=R_5},\right.\nonumber\\
\varepsilon_2 &=& \frac{\partial L}{\partial C}
\left|_{\hat{\phi} =0,R=R_6},\right.\nonumber\\
\varepsilon_3 &=&  \frac{\partial L}{\partial G}
\left|_{\hat{\phi} =0,R=R_7},\right.\nonumber\\
\varepsilon_4 &=&  \frac{\partial L}{\partial \Box R}
\left|_{\hat{\phi} =0,R=R_8},\right.\nonumber\\
\eta &=&  \frac{\partial L}{\partial \Box \hat{\phi} ^2}
\left|_{\hat{\phi} =\varphi_9,R=0}.\right.
      \nonumber
\end{eqnarray}
Conditions (\ref{34}) determine the counterterms to be
\begin{eqnarray}
64\pi^2 \delta \Lambda &=& -64\pi^2\left(\frac{m^2\varphi_0^2}{2}
  +\frac{\lambda\varphi_0^4}{24}\right)\nn\\
 & &+z_0(M_{10},M_{11},\lambda)+(N-1)z_0(M_{20},M_{21},\lambda /3),\nn\\
64\pi^2 \delta \lambda &=&z_1(M_{11},\lambda )+(N-1)z_1(M_{21},\lambda
/3),\nn\\
64\pi^2 \delta
m^2 &=& z(\lambda)+(N-1)z(\lambda /3),\nonumber\\
64\pi^2 \delta \xi &=&z_3(M_{13},\lambda)+(N-1)z_3(M_{23},\lambda
/3),\nn\\
64\pi^2 \delta \varepsilon_0 &=& 2Nm^2 \left(\xi-\frac16\right)
\left(1-\ln\frac{m^2}{\mu^2}\right),      \label{35}\\
64\pi^2 \delta
\varepsilon_1 &=&
-2N\left(\xi-\frac16\right)^2\ln\frac{M_{15}^2}{\mu^2},
\nonumber\\
64\pi^2 \delta \varepsilon_2 &=&
-\frac 1 {60} \ln\frac{M_{16}^2}{\mu^2}
-\frac1 {60}(N-1) \ln\frac{M_{26} ^2}{\mu^2},\nonumber\\
64\pi^2 \delta \varepsilon_3
&=& \frac1 {180} \ln\frac{M_{17}^2}{\mu^2}
  +  \frac1 {180}(N-1) \ln\frac{M_{27}^2}{\mu^2},\nonumber\\
64\pi^2 \delta \varepsilon_4 &=& \frac1 3 N\left(\xi-\frac1
5\right)\ln\frac{M_{18}^2}{\mu^2},\nonumber\\
64\pi^2\delta\eta &=&
\frac{\lambda}{6}\ln\frac{M_{19}^2}{\mu^2}
+\frac{\lambda}{18}(N-1)\ln\frac{M_{29}^2}{\mu^2}\nonumber\\
& &-\frac{\lambda^2\varphi_9^2}{108 M_{19}^2}-\frac{\lambda^2\varphi_9^2
(N-4)}{216 M_{29}^2},\nn
\end{eqnarray}
where we have introduced
\beq
M_{1j}^2&=&m^2+\q R_j+\frac{\lambda} 2 \varphi_j^2,\nn\\
M_{2j}^2&=&m^2+\q R_j+\frac{\lambda} 6 \varphi_j^2,\nn
\eeq
and the functions
\beq
z_0(M_0,M_1,\lambda )&=&
  -\lambda m^2\varphi_0^2
  +\lambda m^2 \varphi_0^2\ln\frac{m^2}{\mu^2}
  +\frac{\lambda^2\varphi_0^4}{4}\ln\frac{M_1^2}{\mu^2}\nonumber\\
  & &-M_0^4\log\frac{M_0^2}{\mu^2}+\frac{3M_0^4}{2}
  -\frac{\lambda^4\varphi_0^4\varphi_1^4}{12M_1^4}
  +\frac{\lambda^3\varphi_0^4\varphi_1^2}{2M_1^2},
\nonumber\\
z_1(M_1,\lambda)&=& -6\lambda ^2\ln\frac{M_1^2}{\mu^2}
+\frac{2\lambda^4\varphi_1^4}{M_1^4}
-\frac{12\varphi_1^2\lambda^3}{M_1^2},\nonumber\\
z(\lambda) &=& 2\lambda
m^2\left(1-\ln\frac{m^2}{\mu^2}\right),\nonumber\\
z_3(M_3,\lambda) &=& -2\lambda
\left(\xi-\frac 1 6\right)  \ln\frac{M_3^2}{\mu^2}
-\frac{2\lambda^2\left(\xi-\frac16\right)\varphi_3^2}{M_3^2},\nonumber
\eeq
which describe essentially the counterterms of a single scalar field.

For the sake of generality, we choosed different values
$\varphi_i,R_i$ for the definitions of the physical coupling constants.
This is due to the fact that, in general, they are measured at different
scales, the behaviour with respect to a change of scale being
determined by the renormalization group equations we shall derive in a
while.

After some calculation one finds the renormalized effective action.
According to equation (\ref{216}), we split it into three parts
\beq
\Gamma_{ren}^{(1)}=\Gamma_{0,ren}^{(1)}
+\Gamma_{2,ren}^{(1)}+\Gamma_{3,ren}^{(1)}.\label{217a}
\eeq
As we have seen in the renormalization, it is useful to introduce
functions describing the contributions coming from the different masses.
Thus, we define
\begin{eqnarray}
\lefteqn{
\gamma_0 (m_1,M_{1i},\lambda) =
m^4\ln\frac{m_1^2}{M_{10}^2}
     +\lambda m^2\varphi_0^2\left(\ln\frac{m^2}{M_{10}^2}+\frac 1
2\right)}\nonumber\\ & &+2m^2\left(\xi-\frac16\right)
R\left(\log\frac{m_1^2}{m^2}-\frac12\right)
+\left(\xi-\frac16\right)^2R^2\left(\ln\frac{m_1^2}{M_{15}^2}-\frac3 2
\right)\nonumber\\ & &-\frac{\lambda^2\varphi_0^4}4
\left[\log\frac{M_{10}^2}{M_{11}^2}-\frac32
-\frac{4(M_{11}^2-m^2)(2M_{11}^2+m^2)}{3M_{11}^4}\right] \label{36}\\ &
&+\left\{\left(\xi-\frac16\right)
R\left[\log\frac{m1^2}{M_{13}^2}-\frac32-\frac{\lambda\varphi_3^2}
{M_{13}^2}\right]
         +m^2\left[\log\frac{m_1^2}{m^2}-\frac1
2\right]\right\}\lambda\hat{\phi}^2\nonumber\\ &
&+\left\{\left(\log\frac{m_1^2}{M_{11}^2}-\frac{25}6\right)
+\frac{4m^2(m^2+M_{11}^2)}{3M_{11}^4}\right\}
\frac{\lambda^2\hat{\phi}^4}4 \nn
\end{eqnarray}
and
\beq
\gamma_2(m_1^2,M_{1j}^2)&=&\frac C {60}\ln\left(\frac{m_1^2}{M_{16}^2}
\right)
-\frac G {180} \ln\left(\frac{m_1^2}{M_{17}^2}\right)-\frac 1 3
\left(\xi -\frac 1 5\right)\Delta R
\ln\left(\frac{m_1^2}{M_{18}^2}\right). \phantom{aaaa}\label{wurst}\eeq
Using this, the renormalized versions of (\ref{213}) and (\ref{214})
read, respectively, \beq
64\pi^2
L_{0,ren}^{(1)}&=&-32\pi^2m^2\varphi_0^2-\frac{8\pi^2\lambda\varphi_0^4}
3 + \gamma_0(m_1^2,M_{1j}^2,\lambda)+(N-1)
\gamma_0(m_2^2,M_{2j}^2,\lambda /3),
    \nn\\
64\pi^2
L_{2,ren}^{(1)}&=&\frac{\lambda}{9}\ln\left(\frac{m_1^2}{m_2^2}\right)
  \left[-2\back\Delta\back+\frac 1 2 \Delta \back^2-\frac 1 {4\back^2}
     \Delta \back^4\right]\label{ren1}\\
  & & -\frac{\lambda}{18}(N-1)
\Delta\back^2\ln\left(\frac{m_2^2}{M_{29}^2}\right)-\frac{\lambda}{6}
  \Delta \back^2\ln\left(\frac{m_2^2}{M_{19}^2}\right)\nn\\
& &-\frac{\lambda}{18}\Delta
\back^2\ln\left(\frac{m_1^2}{m_2^2}\right)\nn\\
& &+\gamma_2(m_1^2,M_{1j}^2)+(N-1)
   \gamma_2(m_2^2,M_{2j}^2).\nn
\eeq
The part $L_{3,ren}$ is directly given by equation (\ref{215}) because
here no renormalization of this term is necessary.

When considering the case of a constant background field, $\back
=const$,
the term $L_{2,ren}^{(1)}$ reduces to the contributions of $\gamma_2$,
furthermore, $L_{3,ren}^{(1)}$ is equal to zero by construction.

Summarizing, we have obtained here the one-loop renormalized effective
Lagrangian in the $O(N)$-model in curved spacetime. The next section
will be devoted to an attempt (using RG) to include higher-loop
effects by improving the effective Lagrangian.

\section{RG improved effective Lagrangian for the O(N)-theory}
\setcounter{equation}{0}
We discuss in this section the RG improvement of the effective
Lagrangian. As is well known, the RG improved effective Lagrangian
sums
all leading logarithms of perturbation theory. We start again
from
the multiplicatively renormalizable Lagrangian (\ref{217}) in curved
spacetime, where curvature invariants are necessary in order to make the
theory
multiplicatively renormalizable. Multiplicative renormalizability
results in the standard RG equation for effective actions (and, here,
also for the effective Lagrangian)
\beq
\left(\mu\frac{\abl}{\abl\mu}+\beta_i\frac{\abl}{\abl\lambda_i}
-\gamma\back\frac{\abl}{\abl\back}\right)L_{eff}(\mu,\lambda_i,\back)=0,
\label{s1}
\eeq
where $\lambda_i=\{\lambda,m^2,\xi,\eta,\Lambda,\epsilon_0,
\epsilon_1,\epsilon_2,\epsilon_3,\epsilon_4\}$ are all the coupling
constants of the theory in the matter sector as well as in the external
gravitational field sector, $\beta_i$ are the corresponding
$\beta$-functions and $\gamma$ is the anomalous dimension of the scalar
field.

Application of the method of characteristics to (\ref{s1})
immediately yields the solution
\beq
L_{eff}(\mu,\lambda_i,\back)=L_{eff}(\mu (t),\lambda_i (t),\back (t)),
\label{s2}
\eeq
where
\beq
\frac{d\lambda_i
(t)}{dt}&=&\beta_i(\lambda_i(t)),\,\,\,\lambda_i(0)=\lambda_i,
\label{renorm}\\
\mu(t)&=&\mu e^t,\,\,\,\back (t)=\back \exp
\left(-\ant_0^{t}dt'\,\gamma (t')\right).\nn
\eeq
The one-loop $\beta$-functions of the theory can be easily found, from
the calculations in the previous section, in the following form:
\beq
\beta_{\lambda}&=&\frac{(N+8)\lambda^2}{3(4\pi)^2},\,\,\,\gamma =0,\nn\\
\beta_{m^2}&=&\frac{(N+2)}{3(4\pi)^2}\lambda m^2,\nn\\
\beta_{\xi}&=&\frac{(N+2)}{3(4\pi)^2}\lambda \q,\nn\\
\beta_{\eta}&=&-\frac{\lambda (N+2)}{36(4\pi)^2}      ,\nn\\
\beta_{\Lambda}&=&\frac{Nm^4}{2(4\pi)^2},\label{s3}\\
\beta_{\epsilon_0}&=&\frac{Nm^2\q}{(4\pi)^2},\nn\\
\beta_{\epsilon_1}&=&\frac{2N\q^2}{(4\pi)^2},\nn\\
\beta_{\epsilon_2}&=&\frac{N}{120(4\pi)^2},\nn\\
\beta_{\epsilon_3}&=&\frac{N}{360(4\pi)^2}.\nn
\eeq
The solution of the RG equations (\ref{renorm}) is easily found to be
\beq
\lambda (t) &=& \lambda\left(1-\frac{(N+8)\lambda t}{3\vp}\right)^{-1},
\nn\\
m^2 (t)&=&m^2\left(1-\frac{(N+8)\lambda t}
{3\vp}\right)^{-\frac{N+2}{N+8}},\nn\\
\xi (t)&=&\frac 1 6 +\q\left(1-\frac{(N+8)\lambda t}
{3\vp}\right)^{-\frac{N+2}{N+8}},\nn\\
\eta (t)&=&\eta +\frac{N+2}{12(N+8)}\ln\left[1-\frac{(N+8)\lambda
t}{3(4\pi)^2}\right] ,\nn\\
\Lambda (t)&=&\Lambda -\frac{3Nm^4}{2\lambda (4-N)}
       \left[\left(1-\frac{(N+8)\lambda
t}{3\vp}\right)^{\frac{4-N}{N+8}} -1\right],\nn\\
\epsilon_0(t)&=&\epsilon_0
-\frac{3Nm^2\q}{\lambda (4-N)}
       \left[\left(1-\frac{(N+8)\lambda
           t}{3\vp}\right)^{\frac{4-N}{N+8}} -1\right],\nn\\
\epsilon_1(t)&=&\epsilon_1
-\frac{3N\q^2}{\lambda (4-N)}
       \left[\left(1-\frac{(N+8)\lambda
           t}{3\vp}\right)^{\frac{4-N}{N+8}} -1\right],\nn\\
\epsilon_2(t)&=&\epsilon_2+\frac{Nt}{120\vp},\nn\\
\epsilon_3(t) &=&\epsilon_3-\frac{Nt}{360 \vp}.\nn
\eeq
Note that, as it follows from those expressions, in the IR-limit ($t\to
-\infty$) ---where the theory is asymptotically free--- the
matter-sector
coupling constants tend to approach their conformally invariant values.

Now, using the classical Lagrangian (\ref{217}) as boundary function,
we
get the RG improved effective Lagrangian in the leading-log
approximation: \beq
L_{eff}&=& \frac 1 2 g^{\mn}\abl_{\mu}\back \abl_{\nu}\back +\eta
(t)\Box \back^2+\lambda(t)\frac{\back^4}{4!}\nn\\
& &+\frac 1 2 (m^2(t)+\xi (t)R)\back^2+\Lambda (t)+\ep_0(t)R+\frac 1 2
\ep_1(t) R^2\nn\\
& &+\ep_2 (t) C+\ep_3 (t)G+\ep_4 (t) \Box R.\label{s5}
\eeq
Here, the question of the choice of $t$ appears. As we see from the
results of the one-loop analysis of the previous section, there are two
effective masses: \beq
m_1^2&=&m^2+\frac{\lambda\back^2} 2 +\q R,\nn\\
m_2^2&=&m^2+\frac{\lambda\back^2} 6 +\q R.\nn
\eeq
Evidently, there is no choice of $t$ which eliminates all
logarithms of the perturbation theory. However, we may, of course,
consider separately the different regions for the parameters of the
theory.

In particular, for large $\back^2$ we have the natural choice
$t=(1/2)\ln(\back^2/\mu^2)$. Then eq.~(\ref{s5}) controls the large
$\back^2$ behaviour of the effective potential. Similarly, for large
values of the mass, $m^2>\back^2$, $m^2>|R|$, we have another choice
$t=(1/2)\ln(m^2/\mu^2)$. Eventually, we do not know the way how to solve
this
problem of writing the RG improved effective potential for the theory
under discussion in a {\it unique} form valid for all regions of
the parameters of the theory.

For the massless theory in flat space, the potential (\ref{s5}) adopts
the Coleman-Weinberg form \cite{colemanweinberg73} (for a discussion of
RG improved effective potentials, see also [50]-[53]).
\nocite{einhornjones83,west83,yamagishi83,sher89}
As
has been recognized recently [54]-[56],
\nocite{kastening92,bandokugomaekawanakano93}
\nocite{fordjonesstephensoneinhorn93}
the RG improved effective potential
for massive theories in flat space includes some term corresponding to
the running of the vacuum energy (the potential at zero $\back^2$). In
the above
approach this term appears as the effective cosmological constant
$\Lambda (t)$, that provides a very natural interpretation of it.

Let us restrict ourselves to the case when $N=1$. Then the RG improved
effective potential is given again by (\ref{s5}) (see also
\cite{elizaldeodintsovunv}), with the natural choice
\beq
t=\frac 1 2 \ln\frac{m^2+\frac{ \lambda\back^2 } 2 +\q
R}{\mu^2}.\label{s6}
\eeq
So, we are able to construct a RG improved potential which has the
same form for the whole parameter region in the case $N=1$.

Another case of particular interest is the massless one. Limiting
ourselves here to linear curvature terms, we get the following RG
improved effective Lagrangian for the $O(N)$-theory
\beq
L_{eff}=\frac 1 2 g^{\mn}\abl_{\mu}\back \abl_{\nu}\back +\eta (t)
\Box\back^2+\frac
1 2 \xi (t) R\back^2+\lambda
(t)\frac{\back^4}{4!}+\Lambda(t)+\epsilon_0(t) R,
\label{s7} \eeq
where $t=(1/2)\ln(\back^2/\mu^2)$. Of course, curvature should be
chosen to be slowly varying. In a similar way, taking into account the
fact
that the choice of $t$ is actually unique, one can get the RG improved
effective Lagrangian for a large variety of massless theories in curved
spacetime \cite{elizaldeodintsov93}.

\section{The back-reaction effect in hyperbolic space\-time}
\setcounter{equation}{0}
As a further application of the results of the previous sections, we
would like here to consider
the back-reaction effect. In order to obtain some concrete information
on the back-reaction, we have to restrict ourselves to some spacetime
with a higher
symmetry.
Note that for time-dependent backgrounds the back-reaction problem is
very difficult to study, even in the case of free fields
\cite{birelldavies82,fischettihartlehu79,hartlehu80,anderson83}, due to
the fact that the effective equations include higher derivative terms as
is clearly seen also from our effective Lagrangian (3.10). That is why,
unfortunately, we were not able to apply our quite general formalism
of the first sections to the back-reaction problem on time-varying
backgrounds (that demands very involved numerical calculations).
Note however that recently
an interesting approach to the back-reaction has been developed
\cite{parkersimon93}, which shows the way to reduce the problem to the
low-derivatives case.
Instead, we limit ourselves to the space of constant curvature.
We choose the spacetime $\reals^2\times H^2/\Gamma$, where
$\Gamma$ is a co-compact discrete group in $PSL(2,\reals)$ containing
only hyperbolic elements \cite{hejhal76}. This choice is interesting
for
the following reasons. First of all, it has been shown that nontrivial
topology might have considerably influenced the early universe and is
in principle observable \cite{ellis71,fagundes93}. From another point of
view, for that
spacetime exact results for the one-loop effective potential are at hand
\cite{bytsenkokirstenodintsov93},
and one is able to see, in which specific range the
effective potential
obtained in this paper is a good approximation to the exact result, and
where it fails to be so.

Let us first start our considerations from the
RG
improved effective potential (\ref{s5}), choosing for simplicity $N=1$
and
$m^2=0$. We are aware of the fact that in this case our
approximation is
not quite good, since we are loosing the part of the whole
information
connected with the non-trivial topology. However, we hope to get
some
hint from the results obtained, because we have included some higher
loop corrections in this approach and, also, below we will present
a similar analysis using the outcome of the mentioned explicit one-loop
calculation
of the effective action on $\reals^2 \times H^2/\Gamma$, thus showing
the influence of the topology in detail. This will also justify our
qualitative result for the purely quantum solution (see (5.8)).

Taking into account the properties of the space under consideration,
namely \beq
R= - \frac{2}{\rho^2}, \ \ \ \ \ \int d^4x \, \sqrt{g} = V(F) \,
\rho^2,
\eeq
we obtain
\beq
\frac{S_{eff}}{V(F)} = \lambda (t) \frac{\back^4\rho^2}{4!} - \xi
(t) \back^2 + \Lambda (t) \rho^2 - 2\epsilon_0(t) +
\frac{2\epsilon_1(t)}{\rho^2},
\label{sef1}
\eeq
where
\bea
t= \frac{1}{2} \log \frac{\lambda \back^2/2- 2(\xi -1/6) \rho^{-
2}}{\mu^2}, && \lambda (t)= \lambda \left( 1 - \frac{3\lambda
t}{(4\pi)^2} \right)^{-1}, \nn \\
\xi (t) = \frac{1}{6} + \left( \xi -  \frac{1}{6} \right) \left( 1
- \frac{3\lambda t}{(4\pi)^2} \right)^{-1/3}, && \epsilon_1(t) =
\epsilon_1 - \frac{1}{\lambda} \left( \xi -  \frac{1}{6} \right)^2
\left[ \left( 1 - \frac{3\lambda t}{(4\pi)^2} \right)^{1/3} - 1
\right]. \nn
\eea
Notice that here the only non-trivial parameter of the metric is
$\rho$. Hence, the effective equations are given by:
\beq
\frac{\partial S_{eff}}{\partial \back} =0, \ \ \ \ \frac{\partial
S_{eff}}{\partial \rho} =0, \ \ \ \ \epsilon_0(t)= \epsilon_0,  \
\ \ \ \Lambda(t)= \Lambda.\label{effe}
\eeq
Then, using $S_{eff}$ (\ref{sef1}), we get
\bea
\back \left[ \frac{1}{6} \lambda (t) \rho^2 \back^2 - 2 \xi (t) +
B(t,\xi,\rho \back) \frac{\lambda}{2} \right] &=& 0, \nn \\
 \frac{1}{12} \lambda (t) \rho^4 \back^4 + 2 \rho^4 \Lambda -
4\epsilon_1(t) + 2(\xi -
1/6)   B(t,\xi, \rho\back)                         &=& 0,
\label{fr1}
\eea
being
\bea
B(t,\xi,\rho \back) &\equiv& \left[  \frac{\lambda \rho^4 \back^4}{4!}
\left( 1- \frac{3\lambda t}{(4\pi)^2} \right)^{-2} - \frac{1}{3}
\left( \xi -  \frac{1}{6} \right)\rho^2 \back^2 \left( 1 -
\frac{3\lambda t}{(4\pi)^2} \right)^{-4/3} \right. \\  && \left. +
\frac{2}{3\lambda } \left(
\xi - \frac{1}{6} \right)^2 \left( 1 - \frac{3\lambda t}{(4\pi)^2}
\right)^{-2/3}  \right]
\, \frac{3\lambda}{(4\pi)^2} \, \left[ \frac{\lambda\rho^2
\back^2}{2} - 2 \left( \xi - \frac{1}{6} \right)
\right]^{-1}.\nn
\eea
Let us now consider some simple, different cases.
\ms

\ni {\bf (1) Case $\back =0$}.  Here we have two possibilities.
\ms

\ni {\it (1a) Subcase $\Lambda \neq 0$}. Then we obtain a
perturbative solution of the form
\bea
\rho &=& \left\{ \frac{2\epsilon_1}{\Lambda} - \frac{2}{\lambda
\Lambda}  \left( \xi -  \frac{1}{6} \right)^2 \left[ \left( 1 -
\frac{3\lambda t}{(4\pi)^2} \right)^{1/3} - 1 \right]\right.\nn\\
& &\left.\qquad +\frac{1}{(4\pi)^2\Lambda} \left( \xi - \frac{1}{6}
\right)^2 \left(
1 - \frac{3\lambda t}{(4\pi)^2} \right)^{-2/3}  \right\}^{1/4} \nn
\\
&\simeq& \left( \frac{2\epsilon_1}{\Lambda} \right)^{1/4} + {\cal O }
(\hbar ). \eea

\ni {\it (1b) Subcase $\Lambda= \epsilon_1= 0$}. In this special
case we obtain a purely quantum solution, by solving
\beq
 1 - \frac{3\lambda t}{(4\pi)^2} -
 \left( 1 - \frac{3\lambda t}{(4\pi)^2} \right)^{2/3} -
\frac{\lambda }{2(4\pi)^2} =0,
\eeq
with $t=(1/2)\ln [-2(\xi-1/6)(\rho\mu)^{-2}]$.
Assuming that $\lambda$ is small, we get (with very good
approximation): $t \simeq -2$, that is
\beq
(\rho\mu)^{-2} \simeq - \frac{1}{2e^2 (\xi - 1/6)},
\eeq
which has sense only for $\xi < 1/6$. Although, as we mentioned, one may
not trust the approximation, we will see that also the one-loop
effective potential leads to the existence of a purely quantum solution.
That we will show below using the one-loop effective potential
calculated with the help of the Selberg trace formula on $\reals ^2
\times H^2/\Gamma$ in \cite{bytsenkokirstenodintsov93}.

\ms
\ni {\bf (2) Case $\back \neq 0$, $\xi = 1/6$}.  Here, assuming
again $\lambda$ small, from eq.~(\ref{fr1}) the following
expressions are easily obtained:
\beq
\lambda \rho^2 \back^2 \simeq 2, \ \ \ \  \lambda \rho^4 \back^4 + 24
\Lambda \rho^4 \simeq 48 \epsilon_1,
\eeq
and combining them, we get
\beq
\rho \simeq \left( \frac{2\epsilon_1 -\dsp \frac{1}{6\lambda}}{\Lambda}
\right)^{1/4}, \ \ \ \ \back^2 \simeq
\sqrt{\frac{2\Lambda}{\lambda}} \, \frac{1}{\sqrt{\epsilon_1
\lambda - 1/12}},
\eeq
which have sense only in the case $\Lambda <0$, $\epsilon_1 < 1/(12
\lambda)$, or else  $\Lambda >0$, but then $\epsilon_1 > 1/(12
\lambda)$. And those are the only possibilities which come out of
the simple cases considered here. All the other cases are
difficult to handle analytically. \newline

To get an idea of the reliability of the approximation, we must first
state the exact result for the $\lambda \varphi^4$-theory
derived in \cite{bytsenkokirstenodintsov93} and then perform a similar
analysis. Without going
into the details of the calculation (the interested reader may consult
\cite{bytsenkokirstenodintsov93}) the effective potential reads
\beq
S_{eff}&=&-\xi \back^2+\frac{2\ep_1}{\rho^2}+\Lambda
\rho^2+\frac{\lambda}{4!}\rho^2\back^4-2\ep_0\nn\\
& &-\frac{\hbar}{128\pi^2}\left\{\frac{25}{12}\lambda^2\back^4\rho^2
+2\lambda\back^2\left[\frac 9 8 -7\xi\right]\right.\label{back1}\\
& &+\frac 4 {\rho^2}\left[3\xi^2-\frac{11}{12}\xi +\frac{79}{960}\right]
     -\frac{8X}{\rho^2}+\frac{16\pi H}{V(\caf)\rho^2}\nn\\
& &-\left[\frac 1 2 \lambda^2\back^4\rho^2-4\lambda \back^2\q+\frac 8
{\rho^2}\left(\q^2+\frac 1 {180}\right)\right]\nn\\
& &\left.\hspace{1cm}\times\left[\ln\left(\frac{\back^2}{M^2}\right)+
\ln\left(1+\frac{2\ep}{\lambda\back^2\rho^2}\right)\right]\right\},\nn
\eeq
where
\beq
X&=&\ant_0^{\infty}dr\,r(r^2+\delta^2)\ln\left(1+\frac{r^2}{\delta^2}
\right)
             (1-\tanh \pi r),\nn\\
H&=&\ant_0^{\infty}dy\,(y^2+2y\delta)\frac{\caz '}
{\caz}\left(y+\delta+\frac 1 2\right),
\eeq
$\caz$ is a Selberg type zeta-function and,
furthermore, $\delta^2=(\xi R +(\lambda/2)\back^2)\rho^2$
and $\ep=\xi-1/8$. It may be shown, that in the limit of small curvature
the resulting equation (\ref{back1}) reduces to the adiabatic
approximation
(\ref{ren1}), apart from finite renormalization terms. From
(\ref{back1}) one easily
finds the effective equation (\ref{effe}) determining the backreaction,
\beq
\lefteqn{
V^{-1}(\caf )\frac{\abl S_{eff}}{\abl \back}=-2\xi\back
+\frac{\lambda} 6 \rho^2\back^3   }\label{back2}\\
& &-\frac{\hbar}{128\pi^2}\left\{\frac{25} 3 \lambda^2\rho^2\back^3+
4\lambda\back\left[\frac 9 8 -7\xi\right]-\frac 8 {\rho^2}
\left(\frac{\abl X}{\abl \back}\right)+\frac{16\pi}{V(\caf
)\rho^2}\left(\frac{\abl H}{\abl \back}\right)\right.\nn\\
& &-\left[2\lambda^2\rho^2\back^3-8\lambda\back\q\right]\left[
\ln\left(\frac{\back^2}{M^2}\right)+\ln\left(1+\frac{2\ep}
{\lambda\back^2\rho^2}\right)\right]\nn\\
& &\left.-\left[\frac 1 2 \lambda^2\back^4\rho^2-4\lambda\back^2\q+\frac
8 {\rho^2}\left(\q^2+\frac 1 {180}\right)\right]\frac{2\lambda
\rho^2\back}
{\lambda\back^2\rho^2+2\ep}\right\}\nn\\
&=&0,\nn
\eeq
\beq
\lefteqn{
V^{-1}(\caf )\frac{\abl S_{eff}}{\abl
\rho}=-\frac{4\ep_1}{\rho^3}-2\rho\Lambda+
\frac{\lambda}{12}\rho\back^4  }\label{back3}\\
& &-\frac{\hbar}{128\pi^2}\left\{\frac{25} 6 \lambda^2\rho\back^4+
-\frac 8 {\rho^3}\left[3\xi^2-\frac{11}{12}\xi+\frac{79}{960}\right]
+\frac{16X}{\rho^3}\right.\nn\\
& &-\frac{32\pi H}{V(\caf )\rho^3}-\frac 8 {\rho^2}
\left(\frac{\abl X}{\abl \rho}\right)+\frac{16\pi}{V(\caf
)\rho^2}\left(\frac{\abl H}{\abl \rho}\right)\nn\\
& &-\left[\lambda^2\rho\back^4-\frac{16}{\rho^3} \left(\q^2+\frac 1
{180}\right) \right] \left[
\ln\left(\frac{\back^2}{M^2}\right)+\ln\left(1+\frac{2\ep}
{\lambda\back^2\rho^2}\right)\right]\nn\\
& &\left.+\left[\frac 1 2 \lambda^2\back^4\rho^2-4\lambda\back^2\q+\frac
8 {\rho^2}\left(\q^2+\frac 1 {180}\right)\right]\frac{4\ep}{\rho}\frac 1
{\lambda\back^2\rho^2+2\ep}\right\}\nn\\
&=&0.\nn
\eeq
In general, it is impossible to solve equation (\ref{back2}),
(\ref{back3}),
explicitly. However, it is easily seen that $\back =0$ is an exact
solution. In this case equation (\ref{back3}) reduces to
\beq
0&=&\frac{4\ep_1}{\rho^4}-2\Lambda\nn\\
& &+\frac{\hbar}{128\pi^2\rho^4}
\left\{-8\left(3\xi
-\frac{11}{12}\xi+\frac{79}{960}\right)+16X-\frac{32\pi H}{V(\caf
)}\right.\nn\\
& &+16\left[\q^2+\frac 1 {180}\right]\left[\ln(2\ep)-\ln(\lambda\rho^2
M^2)\right]\nn\\
& &\left.+16\left[\q^2+\frac 1 {180}\right]\right\}.\label{back4}
\eeq
This can be written as
\beq
\frac 1 {\rho^4}[a+b\ln(\rho^2 M^2)]-c =0,\label{back5}
\eeq
where we have introduced
\beq
a&=&-4\ep_1\nn\\
& &-\frac{\hbar}{128\pi^2}
\left\{-8\left[3\xi
-\frac{11}{12}\xi+\frac{79}{960}\right]+16X-\frac{32\pi H}{V(\caf
)}\right.\nn\\
& &\left.+16\left(\q^2+\frac 1
{180}\right)\left[1+\ln\left(\frac{2\ep}{\lambda}
\right)\right]\right\},\nn\\
b&=&\frac{\hbar}{8\pi^2}\left(\q^2+\frac 1
{180}\right),\nn\\
c&=&2\Lambda .\nn
\eeq
For $\lambda\neq 0$, using the ansatz $\rho =\exp[-a/(2b)]\tilde \rho$
(one has always $b>0$), equation (\ref{back5}) reads
\beq
ce^{-\frac{2a} b}\tilde \rho^4 =\ln (\tilde{\rho}M) .\label{back6}
\eeq
Depending on the values of $a,b,c,$ the number of solutions may be two,
one or none.

If $\Lambda=0$, $\epsilon_1=0$, then one finds the purely quantum
solution $(\rho M)^2=\exp(-a/b)$, where one-loop topological effects are
included exactly and are seen to be of importance. However, it
is of course difficult to find the explicit values of $a$, since $H$ is
very hard to calculate.
Notice that, in principle, in order to be consistent one should
include
into the discussion quantum gravity (for example, in frames of Einstein
gravity). We expect that such an inclusion will not change qualitatively
the solutions of the effective equations.
Note,
finally, that the analysis of the case $\back \neq 0$ is extremely
complicated, due once more to the structure of $H$, and will not be
presented here.

\section{Conclusions}
In new inflationary models the effective cosmological constant is
obtained from an effective potential, which includes quantum corrections
to the classical potential of a scalar field \cite{colemanweinberg73}.
For that reason an intensive effort has been dedicated to the analysis
of the one-loop effective action of a self-interacting scalar field in
curved spacetime. In this paper we have continued this research by
considering a self-interacting $O(N)$ scalar theory in curved spacetime.
First of all we have obtained the derivative expansion of the effective
action
of the theory (sections 2 and 3). It is seen from the analysis,
that two
different effective masses arise, the expansion being consistent if both
effective masses are large as compared to a typical magnitude
of the curvature and to the variation of the background field. Based on
the one-loop analysis we have then discussed the RG improvement in
section 4.
As an application of this study we have considered the
back-reaction
problem in hyperbolic spacetimes of the type $\reals^2\times
H^2/\Gamma$.
The specific way the hyperbolic space is induced as a result of
quantum effects
is clearly seen for some values of the parameters. When comparing the RG
improved result with the exact result available in that case
\cite{bytsenkokirstenodintsov93}, we have seen that topology plays in
fact an
important role in this kind of spacetimes. This, however (as is well
known), may not be seen by using the local adiabatic approximation
scheme.

Let us also say that one may arrive,
qualitatively, to the same conclusions when considering the spacetime
$\reals
\times H^3/\Gamma$, using the results in \cite{cognolakirstenzerbini93}.

As a further application let us mention the symmetry breaking
considerations presented in \cite{cognolakirstenvanzo93} for $N=1$.
Based on our new analysis they can be extended to arbitrary $N$.
In fact, restricting ourselves to $\back =const$, by comparing
with the case $N=1$, the new result is that the quantum corrections to
the mass of the field consist of two pieces, corresponding to the
two effective masses involved. The first piece is the $N=1$ result, the
second piece is $(N-1)$ times the first one with the replacement
$\lambda\to\lambda/3$. Thus the conclusions for special cases, like
maximally symmetric spaces and others (see
\cite{cognolakirstenvanzo93}),
are easily taken over. Especially in the limit $N\to\infty$ the second
mass $m_2^2=m^2+(\xi-1/6)R+(\lambda/6)\back^2$ is the important one and
in this limit the relevant results are found from
\cite{cognolakirstenvanzo93} by putting $\lambda\to\lambda/3$.
Note, however, that these qualitative considerations maybe considerably
changed in a rigorous analysis as by the direct effect of a Goldstone
mode in the spontaneous symmetry breaking phase and also by the possible
appearence of a scalar condensate. These questions will be discussed
elsewhere.

\section*{Acknowledgments}
KK and SDO would like to thank the members of the Department ECM,
Barcelona University, for the kind hospitality.
This work has been supported by the Alexander von Humboldt
Foundation
(Germany), by DGICYT (Spain) and by CIRIT (Generalitat de
Catalunya).

\begin{appendix}
\renewcommand{\theequation}{\mbox{{A}.\arabic{equation}}}

\section{Appendix: Diagonalization of the matrix}
\setcounter{equation}{0}
The aim of this appendix is the diagonalization of the matrix
$U(\back)-(1/6)R$, which is needed for solving
equation (\ref{212}). As given in equation (\ref{23}), we have
\beq
\left(U(\back)-\frac 1 6 R\right)_{ij}&=&d\delta_{ij}+fe_ie_j,\label{A1}
\eeq
with
\beq
d=\frac{U'(\back)}{\back}-\frac 1 6 R,\qquad
f=U''(\back)-\frac{U'(\back)}{\back}.\label{A2}
\eeq
In order to diagonalize the matrix in equation (\ref{A1}), we have to
solve the eigenvalue equation
\beq
U(\back)\vec v ^{(j)}=\lambda_j \vec v ^{(j)},\,\,j=1,...,N.\label{A3}
\eeq
By doing this for $N=2,3$, it is possible to guess the
result for arbitrary $N$. We have found
\beq
\lambda_1=d+f;\,\,\lambda_2=...=\lambda_N=d,\label{A4}
\eeq
with the corresponding eigenfunctions
\beq
v_l^{(1)}=e_l;\,\,
v_1^{(l)}=e_j,\,v_j^{(j)}=-e_1,\,v_k^{(j)}=0,\,\,\,k\neq 1,j.
\label{A5}
\eeq
This determines the matrix $S$, which diagonalizes the matrix $U(\back
)$ by $U_{diag}(\back)=S^{-1}U(\back)S$, to be
\beq
S=\left(\begin{array}{cccccc}
      e_1 & e_2 & e_3 & \cdot & \cdot & e_N \\
      e_2 & -e_1 & 0 & \cdot & \cdot & 0 \\
      e_3 & 0 & -e_1  & \cdot & \cdot & 0 \\
      \cdot & \cdot & \cdot & \cdot & \cdot & \cdot \\
      \cdot & \cdot & \cdot & \cdot & \cdot & \cdot \\
      e_N & 0 & 0 & \cdot & \cdot & -e_1
      \end{array}
   \right),\label{A6}
\eeq
and its inverse
\beq
S^{-1}=\left(\begin{array}{cccccc}
      e_1 & e_2 & e_3 & \cdot & \cdot & e_N \\
      e_2 & -\frac{1-e_2^2}{e_1} & \frac{e_3}{e_1}e_2 & \cdot & \cdot &
 \frac{e_N}{e_1}e_2 \\
      e_3 & \frac{e_2}{e_1}e_3 & -\frac{1-e_3^2}{e_1}  & \cdot & \cdot &
\frac{e_N}{e_1}e_2 \\
      \cdot & \cdot & \cdot & \cdot & \cdot & \cdot \\
      \cdot & \cdot & \cdot & \cdot & \cdot & \cdot \\
      e_N & \frac{e_2}{e_1}e_N & \frac{e_3}{e_1}e_N & \cdot & \cdot &
- \frac{1-e_N^2}{e_1}
      \end{array}
   \right).\label{A7}
\eeq
With these results at hand, equations (\ref{213})-(\ref{215}) are
found from
\beq
tr\left\{\exp\left[-it\left(U(\back)-\frac 1 6 R\right)
\right]a_i\right\}&=&
tr\left\{S^{-1}\exp\left[-it\left(U(\back)-\frac 1 6 R\right)
\right]SS^{-1}a_iS\right\}\nn\\
& &\hspace{-6cm}=
     tr\left\{\left(
    \begin{array}{cccc}
    \exp(-it[d+f]) & 0 & \cdot & 0 \\
       0 & \exp(-itd) & \cdot   & 0 \\
     \cdot & \cdot & \cdot & \cdot \\
     0 & 0 & \cdot & \exp(-itd)
    \end{array}
  \right)
  S^{-1}a_i S\right\}.\label{A8}
\eeq
\end{appendix}

\end{document}